\documentclass{PoS}

\title{International Lattice Data Grid: Turn on, plug in,and download }

\ShortTitle{ILDG}

\author{\speaker{C.M.~Maynard}\\
        EPCC, School of Physics, University of Edinburgh, UK\\
        E-mail: \email{c.maynard@ed.ac.uk}}

\abstract{In the beginning there was the internet, then came the world wide web, and now there 
is the grid. In the future perhaps there will be the cloud. In the age of persistent, 
pervasive, and pandemic networks I review how the lattice QCD community 
embraced the open source paradigm for both code and data whilst adopting the 
emerging grid technologies, and why having your data persistently accessible via 
standardized  protocols and services might be a good idea.}

\FullConference{The XXVII International Symposium on Lattice Field Theory - LAT2009\\
		 July 26-31 2009\\
		 Peking University, Beijing, China}

\begin{document}

\section{A brief history of everything}
The International Lattice Data Grid (ILDG) is a project which has been
running for seven years. The basic idea is to use technology such as
the grid to enable different groups to share their data. It is a
volunteer effort, in that there is no central authority, nor is there
any central funding for effort. There is no central data
repository. Each participating group and individual contributes what
they can in terms of effort, services and data, to make an aggregation
of data accessed by defined services which makes up the ILDG. In this
paper I review the key ideas and explain the rationale behind the
project and motivate why this activity is useful, in addition to our
more usual activities of writing computer code and science
papers. Indeed, a more systematic approach to data management is a
necessary addition! A technical paper on the ILDG services and protocols is
available on the e-print archive~\cite{ILDG}.

In the beginning was the Internet. It was and is a way of connecting
machines together, with multiple routes to different machines. The
basic ideas were developed in the late sixties and early seventies,
and by the nineties a global Internet existed, which continues to grow
to this day. Then came the world wide web, which utilised hyper-text
markup language as a way of sharing information. The world's first web
server was at CERN and went online in 1991. In the late nineties the
``information super-highway'' began to be talked about. In academic
circles, this became known as the grid. In the business world this
became known as web-services. In a direct analogy with the world wide
web, these services and protocols are for sharing data and 
services which may act on data. The future, potential next stage of
evolution is sometimes called cloud computing. Scalable, transparent,
virtual services are provided to users, who access there data via
these services from mobile devices, phones, laptops etc. Everything
lives in the cloud and is operated on by clients.

These networks are persistent, pervasive and pandemic. In an ideal
world so would be our data.  Our data access patterns and methods
should reflect this. How do we access our data? The same way we did a
decade or more ago\footnote{Technologically equivalent with the stone
  age}, with ssl terminal clients (ssh) and copy protocol
(scp). However, our data has changed. We are facing a data
explosion. Data volumes have certainly increased as the size of the
lattice has increased, we now routinely deal in Tera Bytes, and
numbers of configurations belonging to an ensemble have also
increased. We also have to cope with increased data complexity. We
have many, many different ensembles with different quark masses, gauge
couplings and lattice volumes, with many, many different measurements
on each ensemble. Moreover, the ever increasing cost of data
generation and the complexity of the analysis forces groups to join
together to share resources in terms of both computer and man
power. These ``Mega-collaborations'' can be distributed across
continents and time-zones, as can the data. Some tools to help manage
access to the data are becoming a necessity.

How did this activity get stared? The twin philosophies of open source
code and open data started to gain credence around the turn of the
century. The open source code movement was well known to the lattice
community. Many people are familiar with the Linux operating system, and the GNU
compilers, as well as many of the everyday tools such as open
ssl. Would any of the open source ideas be relevant to actual physics
code?  In terms of computation, all the community does the same thing,
{\em viz.} parallel linear solvers. There is only a relatively limited
amount of technical effort available, most of the community is
interested in doing physics rather than assembler coding {\em per se}, so it
does make sense to avoid duplicating work. Many physics codes are now
open source. For example, consider the USQCD~\cite{USQCD} common
programming environment. Development of architecture optimised kernels
are centrally resourced and made available to the community via
standardised interfaces such that legacy codes, the CPS, MILC, and
Chroma, can all call these kernels. The codes are legacy in the sense
of existing before the common programming environment. They represent
many thousands of ``man-hours'' of physics coding, which is now
protected by the development of these kernels.

Open data is a relatively new concept. Many groups had shared data
before, but the MILC collaboration stated to make their data freely
available via the gauge connection~\cite{GC}. This was ground
breaking, in that anyone could use their gauge configurations. Many
people have benefited from their approach and giving away their data
has not harmed MILC's scientific program. Indeed, they have gained
such kudos that other groups have started to copy this strategy.

The emerging grid technologies, such as globus~\cite{globus}, made
distributed computing and in particular distributed data systems a
possibility. Richard Kenway made a proposal that the community should
get together and harness these ideas to make sharing data technically
feasible. At an open meeting at the lattice conference in Boston in
2002~\cite{ILDGBoston2002} this was favourably received and the
International Lattice Data Grid was born!

\section{What is the ILDG?}
There is an organisation, and an aggregated data
repository called the ILDG. Let us consider the organisation. Each
group taking part in the ILDG decides what data is
available to others.  As such, the ILDG has no formal role. Groups
collaborate informally, based around two working groups. One for the
metadata and one for the middleware.  Individual groups were already
starting to build their own, or use existing, data grid
infrastructures. The Adelaide group was developing a
web-portal. Several German groups had combined to make LATFOR, which
had its grid arm, LDG. The Japanese groups had started the Japanese
Lattice Data archive, which became the Japanese lattice data
grid. UKQCD was developing its own grid system, QCDgrid, later
DiGS. The US groups had formed USQCD which had its own data repository
functionality. The middleware for each of these systems was dictated
by national funding considerations. The ILDG is an aggregation of these
middleware systems, and making them interoperable is the key task for
the ILDG. The ILDG data repository can be thought of as a 'grid-of-grids'.

There are three required conditions for the ILDG to exist. Firstly
trust. This is already established in the community. We already know
each other, and review each others work and grant proposals. We may
not always agree on the science, but it is almost inconceivable that
some would claim gauge configurations another group had generated as
their own.  The second condition is altruism. There needs to exist the
political will to make the data available. This is perhaps not so
hard, as we already share code and publish our results and ideas
freely. However, there is effort required collectively to build the
infrastructure and individually to publish, or make available, the
data.  The third condition is reward. How to reward those making the
data available, both individually and as a collaboration. Typically
this is done by users of the data citing a designated paper. This is
the mechanism used to reward those who are involved with machine
building or coding effort and the result should be a highly cited paper.

There are three key ideas or concepts to make this work. Firstly a
standard format for the data. It really doesn't matter what this
format is as long as everyone can read and write data in this
format. One of the first activities of the metadata group was to agree
this format. For configurations this is the SciDAC LIME records, where
the gauge configuration data itself is the NERSC data layout, storing
the full $3\times3$ $SU(3)$ matrix. Secondly, there must be a standard
description of the metadata. This is a semantic description of the
data, which can be processed by an application. A more detailed
discussion of the metadata will follow. Thirdly, standard interfaces to
services are required, so that the same query can be submitted to all
the metadata catalogue services (MDC) and File Catalogue Web services
(FC), what ever underlying middleware system is running them, and critically,
supply an response in an standard format.

\begin{figure}
\begin{center}
\includegraphics[width=.65\textwidth]{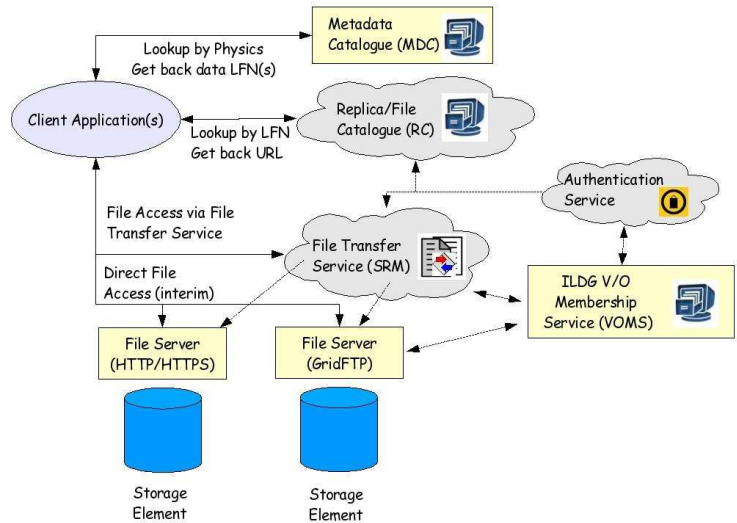} 
\caption{A schematic diagram showing the architecture of the ILDG services}
\label{fig:ILDG-arch}
\end{center}
\end{figure}

A complete a complete description of the architecture can be found on
the ILDG web-pages~\cite{ILDG-web}, is reviewed in~\cite{ILDG} and is shown in
figure~\ref{fig:ILDG-arch}. Access to ILDG services is granted to
users who belong to the ILDG Virtual organisation (VO). To join, a
user needs an X509 certificate from their regional/national academic
certification authority.  The distinguished name (DN) which is part of
the certificate and is the user's identity on the grid is registered
with the ILDG VO, and confirmed by a representative from the regional
grid to which the user belongs. More details on how to do this can be
found on the ILDG web pages~\cite{ILDG-cert}. Who is a member of the ILDG is 
thus controlled by the regional grid organisations.

To access data, a user starts the client application which can submit
physics queries to the MDC. This returns the Logical File Name (LFN)
of the data, this a unique identifyer for a particular datum. The MDCs
are public, and can in principle be read by anyone, but many actually
require authentication, i.e. who is reading the MDC, even if the
information is public. This is controlled via the authentication
service, which talks to the ILDG Virtual organisation using VOMS. 
The client application then submits the LFN to the FC which
returns a URL which encodes the location of the data. A file transfer
is initiated between the storage location chosen by the FC and the location
chosen by the client.

\section{Metadata}
The key concept for the ILDG is metadata. Metadata literally means data about data. This is crucial when sharing and archiving the data. What data is contained in the file? How was it created? What was the physics, the algorithm, computer code and machine? Who created the data? and when? These are all obvious questions about the data. When using data created by another group it is important that this information is recorded in a standard way, so that it can be understood. To motivate the need for a standard further, let us consider the extreme case of no metadata. All the files for the configurations from many ensembles have random strings for names, and all live in the same directory.  This is clearly useless. The obvious thing to do is to organise the files into different directories for each ensemble, and encode some of the metadata into the names of the directories and files. The files code have the position in the Markov chain denoted by the trajectory number. This is constructing a basic scheme or set of rules for describing the data.

This basic and commonly used scheme has ``meaningful'' file names, {\em i.e.} the metadata is encoded in the file names. The directory structure can be used to encode
some structure or hierarchy into the scheme, but it is not completely flexible. The directory structure is a tree, which doesn't map terribly well onto the hierarchies 
inherent in lattice QCD data. For example, if there were ensembles with multiple volumes and quark masses, should the directory structure be \verb-volume/mass- or {\em vice versa}? What happens if data with another flavour of quark is added? It doesn't fit
into the structure, because the scheme is not extensible, {\em i.e.} new data cannot be
added without changing the scheme in such a way that the existing metadata has to be 
modified.

\verb-D52C202K3500U010010_LL3450X_FL3400X_CMesonT00T31- is a UKQCD file name for a meson correlation function. It is an example using the file names to encode the metadata, and is a broken scheme. What does \verb-X- denote? Originally this position in the file name was used to describe the action. $\{W,R,C\}$ meant Wilson, Rotated or (tree-level) Clover.  However, there then turned out to be many sorts of Clover action, depending on the value of the coefficient. So this was denoted with \verb-X- meaning ``none of the above''.
The scheme is broken, there is no where to record the value of another coefficient. To
do so consistently would require a new scheme, and old meson file names would not
fit in this new scheme. Later, UKQCD started doing dynamical fermion simulations, so some metadata for the quark action was recorded in the first part of the name, \verb-D- denotes dynamical. \verb-C202- denotes $c_{SW}=2.02$, but only to three significant figures.
The actual value in the simulation was $c_{SW}=2.0171$. A simple scheme based on meaningful
file names is a sensible first start, but is ultimately inadequate as there is a poor
match between the hierarchical nature of lattice QCD data and the tree structure of
a directory scheme and critically, the lack of extensibility of such a scheme.

What is required from a metadata scheme? There should be sufficient information that
the data can be discovered, accessed or found from the metadata, {\em i.e.}  where is this data. The nature of the data should be encoded in the metadata. This is known as data provenance. What is the data? Can the data be recreated from the metadata? If this
criterion is satisfied then data provenance is assured. However, this is a rather difficult criterion to satisfy in practice. Consider the same ensemble generated on different machines, or with different codes, or even with a different algorithm. What does ``the same'' mean? What level of data provenance is ``good enough''? The scheme should be extensible. New types of data will require new metadata and so the scheme will need to be extended. However, the scheme should be able to be modified in such a way that
the new scheme includes the old scheme. Existing documents which conform to the old
scheme should also conform to the new scheme.

What language should be used to construct such a scheme? Markup languages are an obvious
choice. Markup languages combine text and information about that text. There are different types of markup and a particular language may combine different elements.
Presentational markup is about the format of the text, for instance font type, text size 
or text positioning can be used to allude to relationships between different text fragments. Procedural markup contains text fragments which detail how the text is to be
presented. Tex and Postscript are both procedural markup. Descriptive or semantic markup
contains fragments which label other fragments but no interpretation, presentational or
otherwise is mandated. HTML has both procedural and semantic elements, these pieces are
more formally separated the more recent XHTML language. The semantic extensible markup
language XML was chosen as it is the WC3 open standard~\cite{XML} for a semantic markup language.

The metadata is contained in XML instance documents (IDs) and the rules to which these
IDs must comply are described by an XML schema, called QCDml. There is a complete description of the XML standard at~\cite{XML} but it is useful to consider some of the points raised in the W3-school tutorial on XML~\cite{XML-tut}, which is paraphrased below.
\begin{itemize}
   \item {\em XML was designed to carry data, not to display data.} Incompatible applications can exchange data wrapped in XML. It simplifies data transport and data 
sharing.
   \item {\em XML is just plain text.} It is easy to read and write. User defined tags allow structure to be developed. Lattice QCD data is hierarchical in nature.

   \item {\em XML is designed to be self-descriptive.} It can be read by a person, but 
   see the next point!
   \item {\em XML doesn't DO anything}. The application decides what to do with the 
   data. An application is required to do something with the data.
   \item {\em An XML schema describes the structure of an XML ID.} So an application
   knows how to parse an XML ID.
   \item {\em XML Schemas are the successors of Document Type Definitions (DTDs).} They
   have richer functionality, support data types and namespaces, are extensible and are
   themselves an XML ID.
\end{itemize}
A description of QCDml itself is given in~\cite{Maynard:2004wg} and how to markup 
configurations in QCDml is described in~\cite{Coddington:2007gz}.

\section{QCDml now and in the future}
The biggest challenge for metadata usage is metadata capture. The moment of data creation is when all the metadata is known. Capturing this information is difficult because the goal of the simulation is to generate the data, not the metadata. At the time of writing, there are no codes writing QCDml as the data is generated. This is partly a consequence of XML parsers being mostly written for Java. The qdp++/chroma code~\cite{Edwards:2004sx} suite does read and write XML at run-time using the libxml2 library and could, in principle be adapted to write QCDml. However, for most codes, it would require a significant coding effort to adapt and critically maintain, a QCDml parsing interface, when the same effort could be deployed on physics functionality or code performance. Most applications write some metadata when the data is written, possibly even a simple plain text file. This then has to be converted to QCDml at a later point. Whilst this is better than not recording metadata, post-processing data is laborious and often {\em ad-hoc}, in that it doesn't happen automatically. It usually requires the intervention of an individual, when they are not too busy with something else. This means there is often a significant delay which results in an onerous task on the individual concerned\footnote{The author personal experience of this.} or worse result in a loss of the metadata.

QCDml has a rich description of the action used to generate an ensemble, and some information regarding who, on which machine, with which code, but doesn't contain
a lot of information regarding the algorithm. QCDml doesn't ensure data provenance, certainly, the data cannot be recreated from the metadata alone. However, the data can be discovered from the metadata by asking physics questions, so the first requirement for the metadata is satisfied even if the second isn't. 

There are no plans to significantly alter QCDml, and the situation with metadata capture
is also unlikely to change. However, a possible solution to these difficulties can be
found by adopting some workflow tools. Besides automating the process of data generation
workflow tools record all information about the stages of the calculation automatically.
This helps with data provenance, and with metadata capture. This doesn't solve both problems, but provides a basis for the automation of metadata capture and increases the
amount of data provenance. The reader is referred to the work of Simone's group at Fermilab~\cite{Piccoli:2008zz}.

Several groups have expressed interest in using the ILDG to transport or share other types of lattice data
such as propagators. The ILDG working groups have adopted the USQCD propagator formats, with an extension
proposed by LDG for twisted mass propagators. Metadata for this type of data has yet to be agreed, but
a very lightweight scheme, {\em i.e.} one with minimal content, is desired by the interested parties. The ILDG
hope to adopt such a scheme in the near future.

\section{Conclusions}
In the last decade Lattice QCD data has undergone a dramatic growth in complexity. The rise of the Mega-collaboration to cope with the increased amount of work required to complete calculations
has also lead to the data being geographically distributed. Managing the data ``by hand'' is no
longer scalable. Tools are required to automate the logistics of data management and to provide
data provenance. Regional grids were developed in response to this need and the ILDG sits on top of these
architectures. Remote data can be accessed with local tools, but data standards are required, which in
turn enforces {\em data discipline} to curate the data. This is probably a good discipline to have.
It is hoped the ILDG can encourage scientific innovation as making data readily accessible so that
good ideas are not thwarted by a lack of opportunity to exploit them. Many of the large collaborations
are making their data available in this way, which can only be of benefit to the whole community, but
this altruistic effort from both collaborations and individuals should be recognised.

\end{document}